\documentclass[twocolumn, tighten]{aastex63} %
\usepackage{multirow,amsmath,graphicx,xspace,mathrsfs,dutchcal}
\newcommand{\A}{$\mathbb{A}$}
\newcommand{\Mc}{$\mathrm{Mx\ cm^{-2}}$}
\usepackage{enumitem}
\usepackage{lineno}

\begin{document}
\title{\textbf{Asymmetric Distribution of the Solar Photospheric Magnetic Field Values}}
\correspondingauthor{Jing-Chen Xu}
\email{jcxu@ynao.ac.cn}

\author[0000-0002-8947-547X]{Jing-Chen Xu}
\affiliation{Yunnan Observatories, Chinese Academy of Sciences, Kunming 650011, China}
\affiliation{State Key Laboratory of Space Weather, Chinese Academy of Sciences, Beijing 100190, China}
\author[0000-0003-3944-9916]{Ke-Jun Li}
\author[0000-0003-2921-1509]{Peng-Xin Gao}
\affiliation{Yunnan Observatories, Chinese Academy of Sciences, Kunming 650011, China}
\affiliation{Center for Astronomical Mega-Science, Chinese Academy of Sciences, Beijing 100012, China}


\begin{abstract}
Understanding the characteristics of the solar magnetic field is essential for interpreting solar activities and dynamo. In this research, we investigated the asymmetric distribution of the solar photospheric magnetic field values, using synoptic charts constructed from space-borne high-resolution magnetograms. It is demonstrated that the Lorentzian function describes the distribution of magnetic field values in the synoptic charts much better than the Gaussian function, and this should reflect the gradual decay process from strong to weak magnetic fields. The asymmetry values are calculated under several circumstances, and the results generally show two periodicities related to the variation of the solar B$_0$ angle and the solar cycle, respectively. We argue that it is the small-scale magnetic fields, the inclination of the solar axis, the emergence and evolution of magnetic flux, and the polar fields that are responsible for the features of asymmetry values. We further determined the polar field reversal time of solar cycles 23 and 24 with the flip of asymmetry values. Specifically, for cycle 24, we assert that the polar polarities of both hemispheres reversed at the same time --- in March 2014; as to cycle 23, the reversal time of the S-hemisphere is March 2001, while the determination of the N-hemisphere is hampered by missing data.
\end{abstract}
\keywords{Solar magnetic fields --- Solar photosphere -- Solar activity}

\section{Introduction} \label{sec:intro}

Magnetic field couples the solar interior, the solar atmosphere, throughout the heliosphere; it is responsible for the many solar activities, heating of the chromosphere and corona, and acceleration of the solar wind~\citep[e.g.,][]{Wiegelmann2014, Komm2015, Cheung2017}; besides, its characteristics are essential for interpreting the solar dynamo and solar cycle~\citep{Brun2017, Charbonneau2020}. Therefore, understanding the spatial and temporal evolution of the solar magnetic field is of vital importance.
Hitherto, only the photospheric magnetic field can be measured in detail routinely with high-resolution space-borne instruments~\citep{Lagg2017}, e.g., the Michelson Doppler Imager onboard the Solar and Heliospheric Observatory (MDI/SOHO; \citealt{Scherrer1995}) and the Helioseismic and Magnetic Imager onboard the Solar Dynamics Observatory (HMI/SDO; \citealt{Schou2011, Scherrer2011, Hoeksema2014}). The interpretation and modeling of magnetic field on other parts of the solar atmosphere are mainly dependent on the photospheric measurements~\citep{Lagg2017}. In this study, we will investigate the asymmetric distribution of the solar photospheric magnetic field values.

The solar surface is overspread with ubiquitous magnetic fields coming from the quiet Sun and active regions~\citep{Sheeley1966, Harvey1971, Jin2011}. The quiet Sun photospheric magnetic field is usually divided into kilogauss magnetic network and internetwork magnetic field with weaker strength; the former is situated in the profiles of supergranular cells where the horizontal flows become downdrafts, and the latter is small-scale flux concentrations scattered on the solar surface \citep{BellotRubio2019}.
The distribution of the magnetic field values is a fundamental property of the Sun.
\citet{Getachew2019a, Getachew2019} studied the asymmetry of solar weak photospheric magnetic field values, using synoptic chart data from several sources, including the Wilcox Solar Observatory, Mount Wilson Observatory, Kitt Peak Vacuum Telescope, SOHO/MDI, SOLIS/VSM, and SDO/HMI. In their study, `asymmetry' is defined as the nonzero field value of the fitted Gaussian distribution maximum. They fitted the distribution with a parameterized Gaussian function, and then the asymmetry is revealed, mainly when the spatial resolution of the synoptic chart is considerably reduced, such as $360\times180$ pixels or even $120\times48$ pixels which are much lower than the original resolution (e.g., $3600\times1440$ pixels for HMI). They showed that the lower the resolution, the more prominent the calculated asymmetry; according to which, they inferred that it is the supergranulation that brings about the asymmetry.

There are several shortcomings in estimating the asymmetry with highly resolution-reduced synoptic charts.
First, high-spatial-resolution magnetograms and synoptic charts resolve quiet Sun small-scale magnetic fields much better, such as internetwork, network, etc. There is no reason to give up the most essential advantage of observation, i.e., high-resolution. For a chart of 3600$\times$1440\ (1080) pixels, the spatial resolution is about $1.68''\times1.33''\ (1.78'')$ (the physical spatial resolution on the solar surface is in the magnitude of ${\sim}10^3$ km). When the spatial resolution decreases to around ${\sim}3''$, most of the magnetic features are canceled and the magnetograms look very vague \citep{BellotRubio2019}. It will be unable to resolve many of the adjacent magnetic elements with opposite polarities.
For charts of $120\times48$ pixels, the spatial resolution is about $50.3''\times40.0''$, which is much larger than ${\sim}3''\times3''$.
Second, the number of pixels in the distribution decreases substantially when a chart is downsized, which affects the fitting result. From $3600\times1440$ to $120\times48$ pixels, the number of pixels decreases by nearly 3 orders of magnitude.
In the distribution of a downsized chart, the probability of adjacent bins vary drastically (e.g., Fig. 2 in \citealt{Getachew2019a}). The parameters of the fitted Gaussian model are unlikely reliable in judging the asymmetry.
Last, the noise of measurements (a few Gauss) should be taken into consideration, because they dominate the location parameter of a Gaussian function during the fitting of magnetic field values, especially when the distribution is narrow.

In this study, we show that the distribution of solar photospheric magnetic field values is much better fitted with the Lorentzian function, rather than the Gaussian function. The asymmetry is clearly revealed after merely excluding the noise. We also uncovered the temporal characteristic of the asymmetry variation. This paper is organized as follows.
In Section 2, we introduce the synoptic chart data. Section 3 introduces the method and Section 4 the analyses and results. In the final Section 5, we conclude the paper and discuss the results.


\section{Data: Synoptic Charts}\label{sec:data}

Due to the substantial instrumental differences between various ground and space-based measurements, we only make use of MDI/SOHO and HMI/SDO high-resolution synoptic charts that are coming from space-borne observations. The HMI started observation with similar yet upgraded instruments when the MDI was reaching its end of life, i.e., the HMI is a successor to the MDI~\citep{Scherrer2011}. Synoptic charts from both of them are constructed from zero-level offset corrected magnetograms~\citep{Liu2004}. The HMI (MDI) charts are 3600 points in Carrington longitude by 1440 (1080) points equally spaced in sine latitude. The inferred radial synoptic charts provide the imputed radial component of the magnetic field over the entire solar disk. Each pixel on a synoptic chart is generated using near-central-meridian data from 20 magnetograms.
A detailed description and comparison of the charts from the two instruments can be found in \citet{Liu2012}. In this study, we used synoptic charts of MDI span from Carrington Rotation (CR) 1909 to CR-2096 (from May 6, 1996 to Dec 10, 2010; 188 charts in total), and that of HMI from CR-2097 to CR-2232 (from May 6, 2010 to Jun 18, 2020; 136 charts in total). In all, there are 324 synoptic charts, spanning 24 years which covers more than two solar cycles.

\begin{figure*}
\centering
\includegraphics[width=\textwidth]{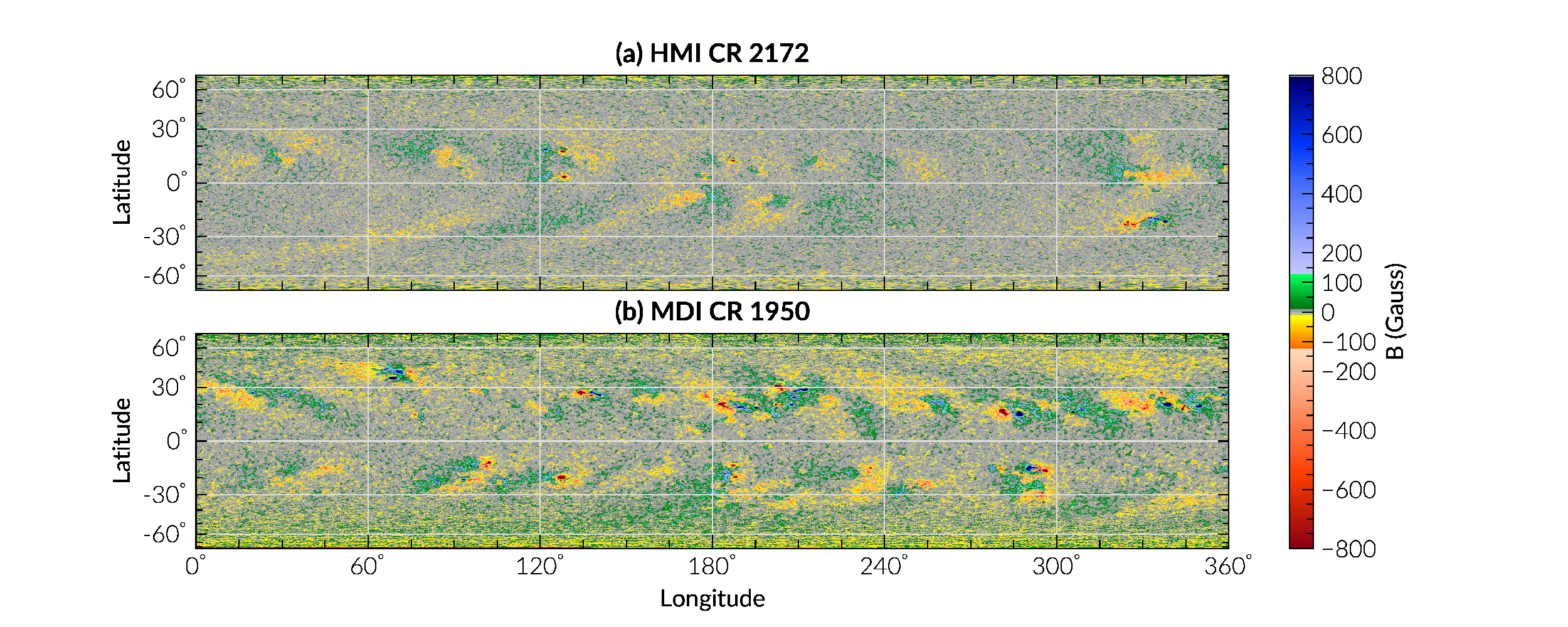}
\caption{Example of synoptic charts from HMI and MDI: (a) HMI CR-2172 (2015/12/25--2016/01/21), and (b) MDI CR-1950 (1999/05/28--1999/06/24). The two subfigures share the same colorbar. The magnetic field saturates at $\pm$800 \Mc. The resolution of HMI (MDI) is 3600$\times$1440 (1080) pixels. The field values measured by MDI is greater than that of HMI by a factor of 1.40, and that is why the magnetic fields in the lower panel seems stronger.}
\label{fig:syn_chart}
\end{figure*}

Examples of HMI and MDI synoptic charts are depicted in Figure~\ref{fig:syn_chart}. In the Figure, the CR number of the HMI chart shown here (CR-2172) is the same as in \citet{Getachew2019a}, and the CR number (CR-1950) of MDI is chosen so that it is roughly amid solar maximum and minimum (similar to the case of CR-2172). The two subfigures share the same colorbar, and the colormap is \texttt{hmimag} from the \texttt{SunPy}~\citep{Community2020} open-source software.
The magnetic field values measured by MDI is greater than that of HMI by a factor of 1.40~\citep{Liu2012}, and that is why the magnetic fields in the lower panel of Figure~\ref{fig:syn_chart} seems stronger.

\section{Method}\label{sec:metho}

To estimate the asymmetry of solar photospheric magnetic fields (denoted as `\A' afterward), we model the magnetic field value distribution of each synoptic chart with a three-parameter Lorentzian function (or Cauchy-Lorentz distribution). The Lorentzian function is written as \[f(x; x_0,\gamma,I) = \dfrac{I \gamma^{2}}{\gamma^{2} + \left(x - x_{0}\right)^{2}}\] where $x_0$ is the location parameter, specifying the location where the distribution reaches the peak. It is interpreted as the solar photospheric magnetic asymmetry \A\ in this study. Besides, $I$ is the height of the peak, and $\gamma$ is half of the full width at half maximum (FWHM). Since only in the special case when $I=1/(\pi \gamma)$, the integration of this function equals to 1, the Lorentzian function is normally not a probability density function.
We use the \texttt{astropy.modeling} package to do the fitting. The model parameters and their uncertainties are estimated with the Levenberg-Marquardt algorithm and the least-squares statistic. The significance of an obtained asymmetric value \A\  is quantified with the same method as in \citet{Getachew2019a} (with different degrees of freedom).

\begin{figure}
\centering
\includegraphics[width=0.4\textwidth]{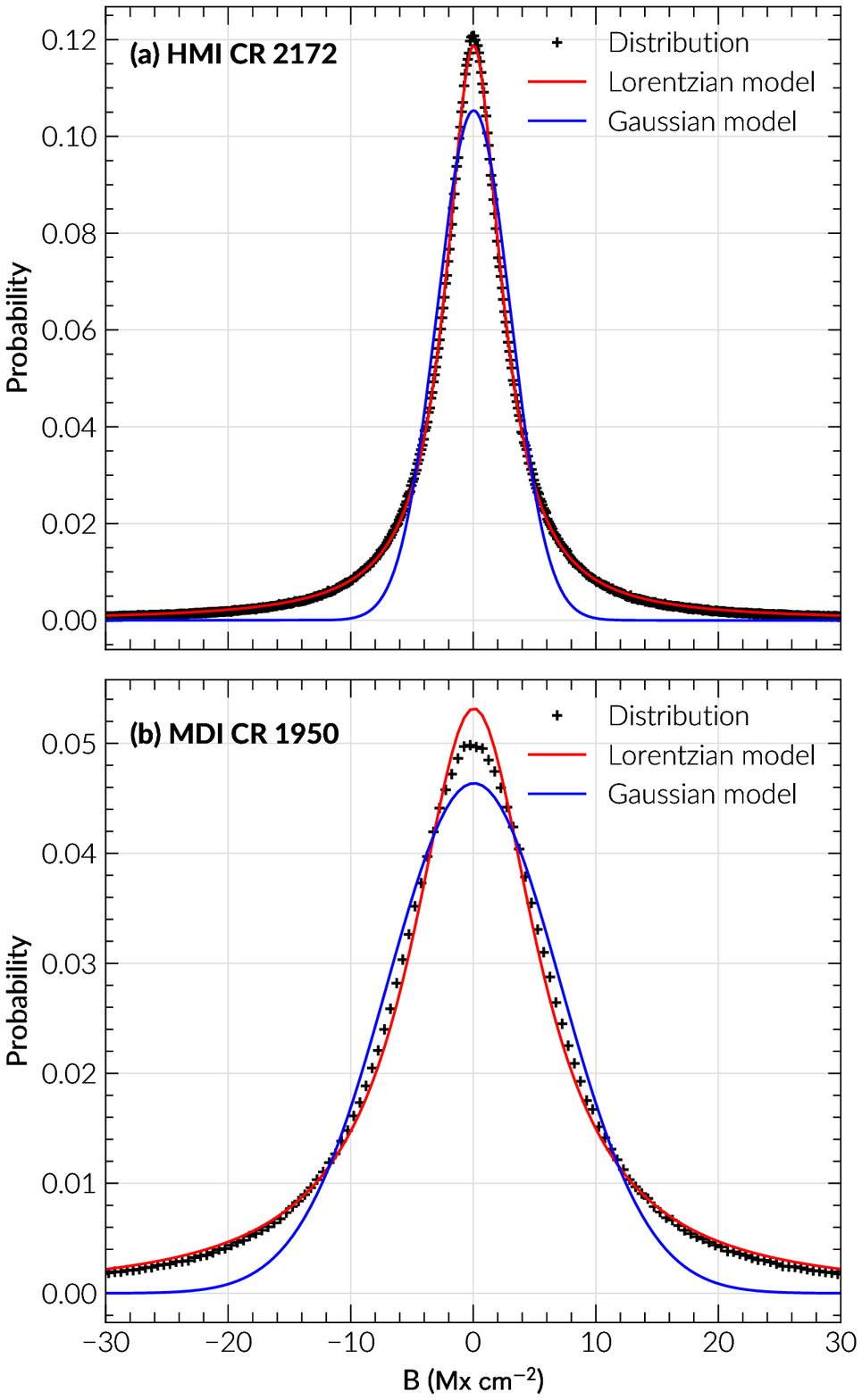}
\caption{Typical examples of the distribution and modeling of magnetic field values in synoptic charts: (a) HMI CR-2172 and (b) MDI CR-1950. The original distribution is plotted with black `+' signs. The distributions are fitted with the Lorentzian model (red line) and the Gaussian model (blue line), respectively.}
\label{fig:model}
\end{figure}

Figure~\ref{fig:model} shows the distributions and model fittings for the two exemplary synoptic charts in Figure~\ref{fig:syn_chart}. Specifically, the magnetic field values in the synoptic charts of HMI CR-2172 and MDI CR-1950 are fitted with both the Lorentzian model and Gaussian model, respectively. The Lorentzian function is an example of a fat-tailed distribution that exhibits a large kurtosis relative to that of a Gaussian function. The result indicates that the Lorentzian model is much more suitable than the Gaussian model in describing the distribution of magnetic field values, both at the center and the tails. It should reflect the property of the evolution of magnetic fields from strong magnetic fields in active regions decaying into weaker fields.

As exhibited in the modeled Gaussian distribution in Figure~\ref{fig:model}, the major part of the distribution is concentrated within about $\pm5$ \Mc\ (for HMI) or $\pm10$ \Mc\ (for MDI), and it plunges to zero in the tails very quickly. It is obvious that the Gaussian model describes the distributions poorly; especially, it seriously understates the tails of the distribution of magnetic field values.
For the HMI and MDI line-of-sight magnetic field synoptic charts, the noise levels are 2.3 and 5.0 \Mc, respectively, and they are quite uniform over the entire chart \citep{Liu2012}.

The magnetic field values larger than tens of Gauss are totally ignored.  Thus the asymmetry is almost completely dominated by the very weak magnetic fields mixed with indistinguishable noise.
Therefore, not surprisingly, the calculated asymmetry with the Gaussian model is not dependent on the upper range of magnetic field values from 10 to hundreds Gauss~\citep{Liu2004, Getachew2019a}.  In this study, in the process of estimating the \A, we will exclude the very weak field values which are unresolvable from noise.

\section{Result}

In this section, first, we calculate the asymmetry values \A\ of each synoptic chart in three different cases. Then, we calculate \A\ after excluding polar regions and high-latitude areas; the temporal characteristics of the obtained \A\ are explored. In addition, we attempt to determine the reversal times of polar field polarities of solar cycles 23 and 24 from the reversal of  \A.

\subsection{Asymmetry values: three cases} \label{subsec:3cases}

We model the distribution of magnetic field values of each synoptic chart with a Lorentzian function to calculate the asymmetry \A\, and estimate the significance in the meanwhile. Here, only the original high-resolution charts of HMI (3600$\times$1440 pixels) and MDI (3600$\times$1080 pixels) are used.  For each chart, the \A\ is calculated in three cases:
\begin{enumerate}
\item model the original distribution (case-1: orig);
\item model the distribution after excluding magnetic field values within range $-5.4$ to 5.4 G (case-2: $|\mathrm{B}|\geqslant 5.4$ G);
\item model the distribution after excluding magnetic field values within range $-10.4$ to 10.4 G (case-3: $|\mathrm{B}|\geqslant 10.4$ G).
\end{enumerate}
In the latter two cases, the noise and very weak fields are excluded. Here, the threshold values 5.4 G and 10.4 G are chosen because they are the average FWHM of HMI and MDI, respectively. In performing the fitting, we limit the range of magnetic field values within $-$600 to 600 G.

The significance of the obtained \A\ values is estimated via the same method as in \cite{Getachew2019a}, i.e., the Student's $t$-tests. To be more specific, we compare the calculated $\hat{t}={Z}/{s}$ value with the pre-calculated statistic $t_{0.01,n-3}$; here, $Z$ is the mean value of the fitted location parameter $x_0$ (i.e., the asymmetry value \A), $s$ is the standard error of $Z$, and $n$ is the degree of freedom.
If $\hat{t}>t_{0.01,n-3}$, the corresponding \A\ is considered to be statistically significant at the 99\% confidence level. The calculation is done with the \texttt{stats.t} module in the open-source \texttt{SciPy} package.

\begin{figure} %
\centering
\includegraphics[width=0.48\textwidth]{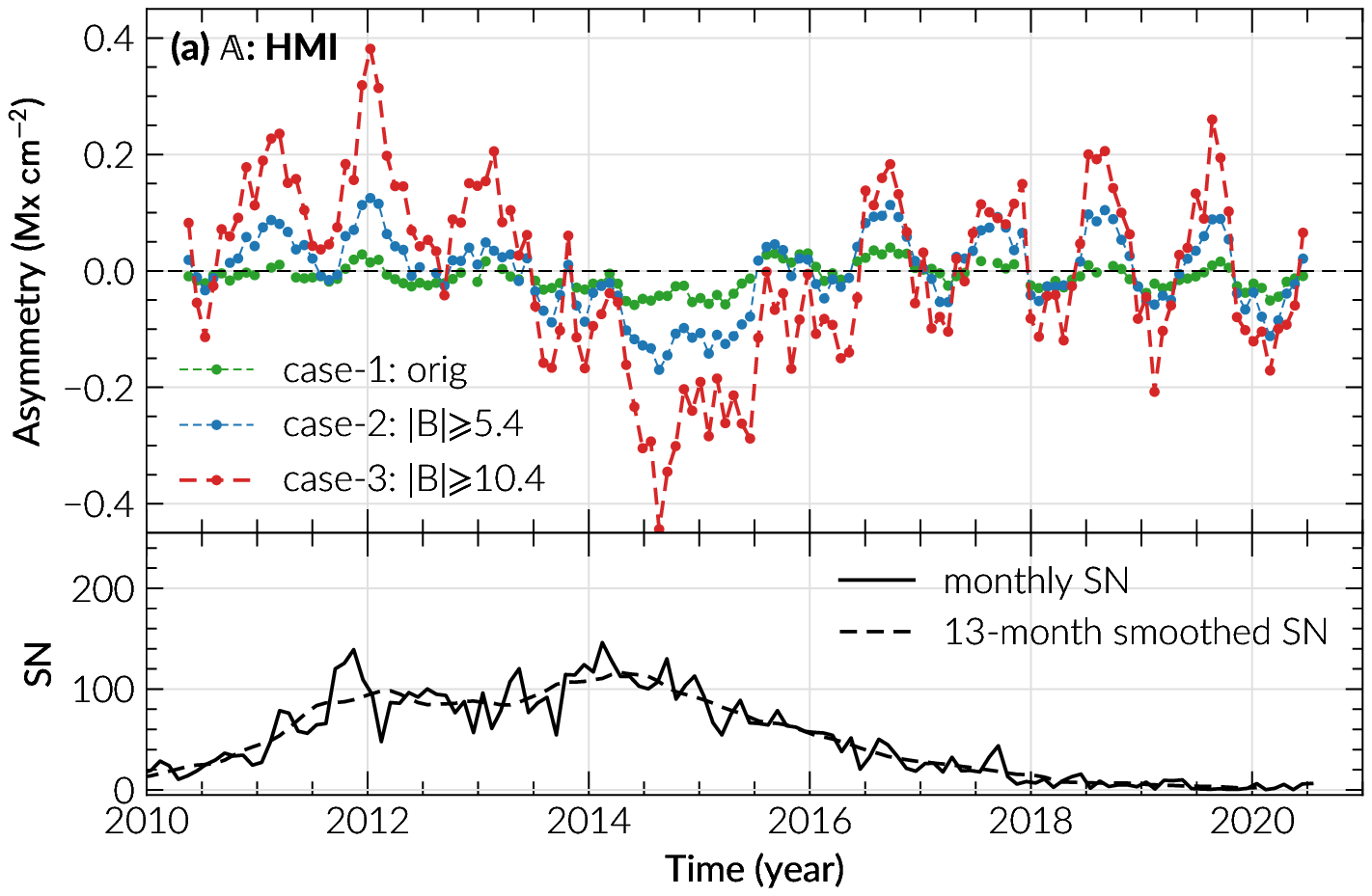}
\includegraphics[width=0.48\textwidth]{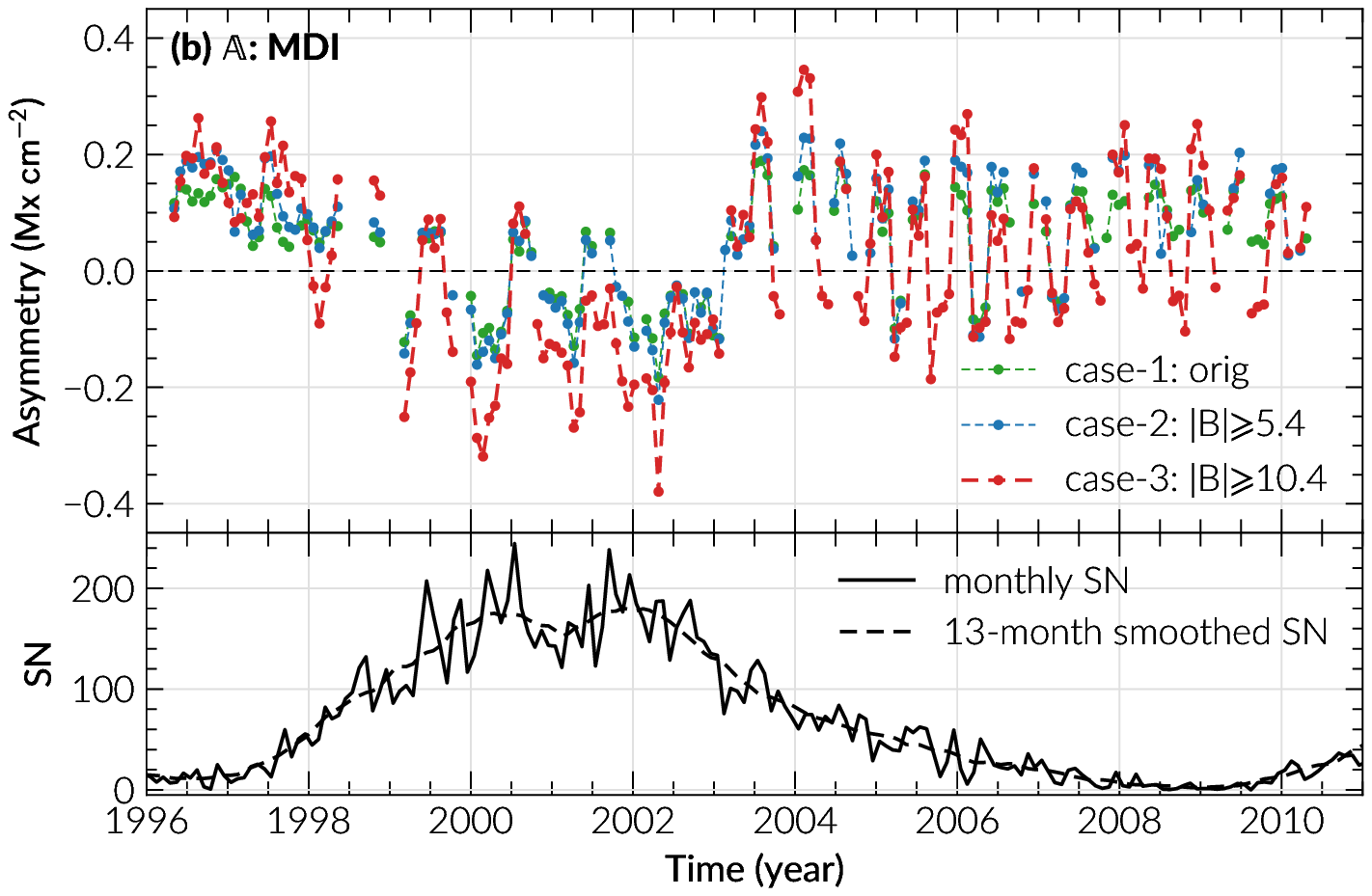}
\caption{Asymmetry values \A\ obtained from (a) HMI and (b) MDI synoptic charts in three cases (see text), as well as a comparison with the sunspot numbers. Only values that are statistically significant at the 95\% confidence level are plotted. The dots are not connected when there are insignificant values. Note that the time spans of the two subfigures are different. The range of the y-axis in the upper panels of the two subfigures are the same.}
\label{fig:asymTwo}
\end{figure}

The resulted \A\ values from the HMI charts are shown in Figure~\ref{fig:asymTwo} (a).
For the three cases for HMI, the counts of significant \A\ values are 124, 133, and 136 (136 in total), respectively, which means that 91.2\%, 97.8\%, and 100\% of the asymmetry values are statistically significant. Compared with \citet{Getachew2019a}, the obtained \A\ values are more noteworthy, and the high percentages of significant values here is the result of a better fitting with the Lorentzian function. When the fitting is performed on a resolution-reduced map, the frequency (or probability) vary drastically between adjacent bins in the distribution (besides the shortcoming of underestimating the fat-tail), and thus the uncertainty is large, which probably leads to a much larger portion of insignificant \A\ (see the right two columns of figure 2 in \citealt{Getachew2019a}).

In performing the modeling for each chart, in the latter two cases (case-2 and case-3) , the amplitudes of the resulted model (i.e., the `$I$' in the Lorentzian function) are usually different from that in case-1. Thus, we also perform the modeling in case-2 and case-3 with the amplitude being equal to the respective amplitude in case-1.
It turns out that fixing the amplitude or not only affect the results marginally.
In addition, changing the bin size from 0.1 to 0.5 \Mc\ or changing the magnetic field values ranges from $\pm$100 to $\pm$1500 \Mc\ affect the asymmetry marginally (the maximum difference of \A\ is less than $4\times10^{-3}$). In case-2 and case-3, \A\ is affected by the threshold values; specifically, it gradually becomes larger with the increase of threshold value from 0 to 10.4 \Mc.

With the same procedure, the \A\ values calculated from the MDI charts are plotted in Figure~\ref{fig:asymTwo} (b). For the three cases, the counts of significant values of \A\ are 139, 143, and 166 (188 in total), respectively, which means that 73.9\%, 76.1\%, and 88.3\% of the asymmetry values are statistically significant. 
The obtained \A\ values in case-3 are in the same order of magnitude as that obtained with a resolution as low as 72$\times$30 pixels in \citet{Getachew2019a}.

Figure~\ref{fig:asymTwo} manifests that \A\ values increase systematically from case-1 to case-3, which is more obviously shown in the results of HMI (panel a). Both of the panels indicate that the \A\ is modulated by the solar cycle. In case-2 and case-3 in panel (a), the \A\ values present opposite relations with the two peaks of the double-peaked maxima of the solar cycle. Specifically, it is clearly seen from \A\ of HMI (red or blue dots connected by dashed lines) that at the former peak during the ascending phase of solar cycle 24, the \A\ surges and stays positive, while at the latter peak, the \A\ plunges and remains negative.
This regularity seems not true for \A\ of MDI at the former peak of solar cycle 23 though, which is possibly due to the data quality of the early observations.
Besides, the \A\ values seem to oscillate ups and downs periodically.

\subsection{Periodicity of the asymmetry values}  \label{subsec:wavelet}

\begin{figure}
\centering
\includegraphics[width=0.48\textwidth]{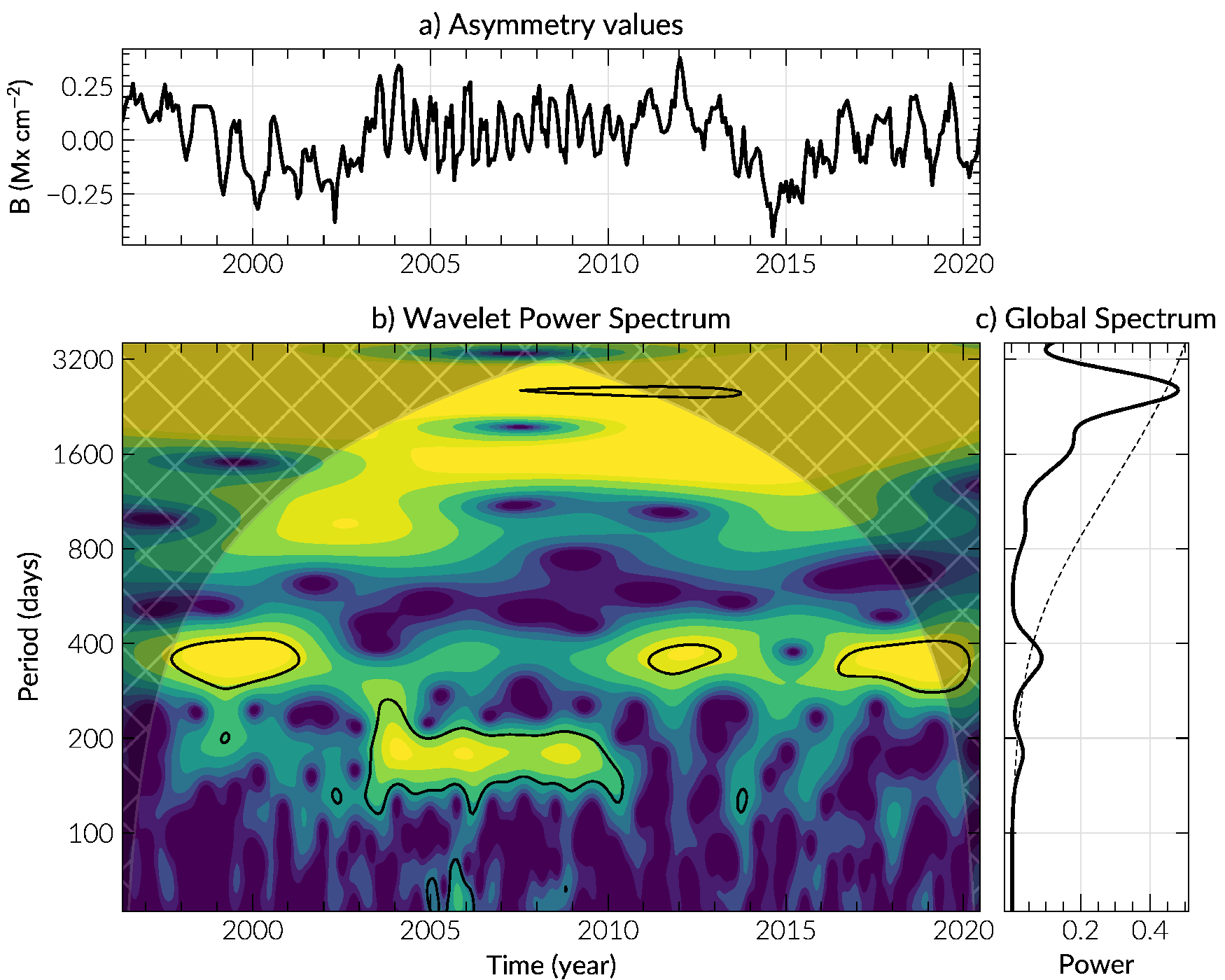}
\caption{Wavelet power spectra of the asymmetry values. (a) The asymmetry values; HMI results are used during the overlapped time. (b) The wavelet power spectrum. The black solid contours indicate the 95\% confidence level, and the hatched area indicates the cone of influence where edge effects become important. (c) The global wavelet power spectrum (solid line) and the 95\% confidence level (dashed line). In (b) and (c) the y axis is logarithmic.}\label{fig:wave}
\end{figure}

The variation of the asymmetry values from both HMI and MDI suggests that there are periodicities in them, especially in \A\ of case-2 and case-3. Therefore, we join the \A\ values from the two instruments of case-3 to form a continuous time series and perform the wavelet analysis; the power spectra are displayed in Figure~\ref{fig:wave}. The wavelet power spectrum in Figure~\ref{fig:wave} (b) reveals that areas above the 95\% confidence level are distributed in three period bands. The global wavelet power spectrum in Figure~\ref{fig:wave} (c) indicates that the three periods are: 178.1 days ($\sim$0.5 year), 361.1 days ($\sim$1.0 year), and 2441.2 days ($\sim$6.7 years), respectively. All of them are statistically significant at the 95\% confidence level. The $\sim$6.7-year period is affected by the cone of influence, which is due to the limited length of \A\ time series.

These obvious characteristics are not revealed in previous studies.
We speculate that the $\sim$0.5 and $\sim$1 year periods are caused by the annual variation of the B$_0$ angle, i.e., the inclination of the solar rotational axis with respect to the ecliptic plane, or the heliographic latitude of the central point of the solar disk.
The $\sim$6.7-year period indicates that the asymmetry is probably related to the solar activity cycle.

\subsection{Asymmetry values: hemispheres and low latitudes}
\label{subsec:FNS}

\begin{figure*}
\centering
\includegraphics[width=0.8\textwidth]{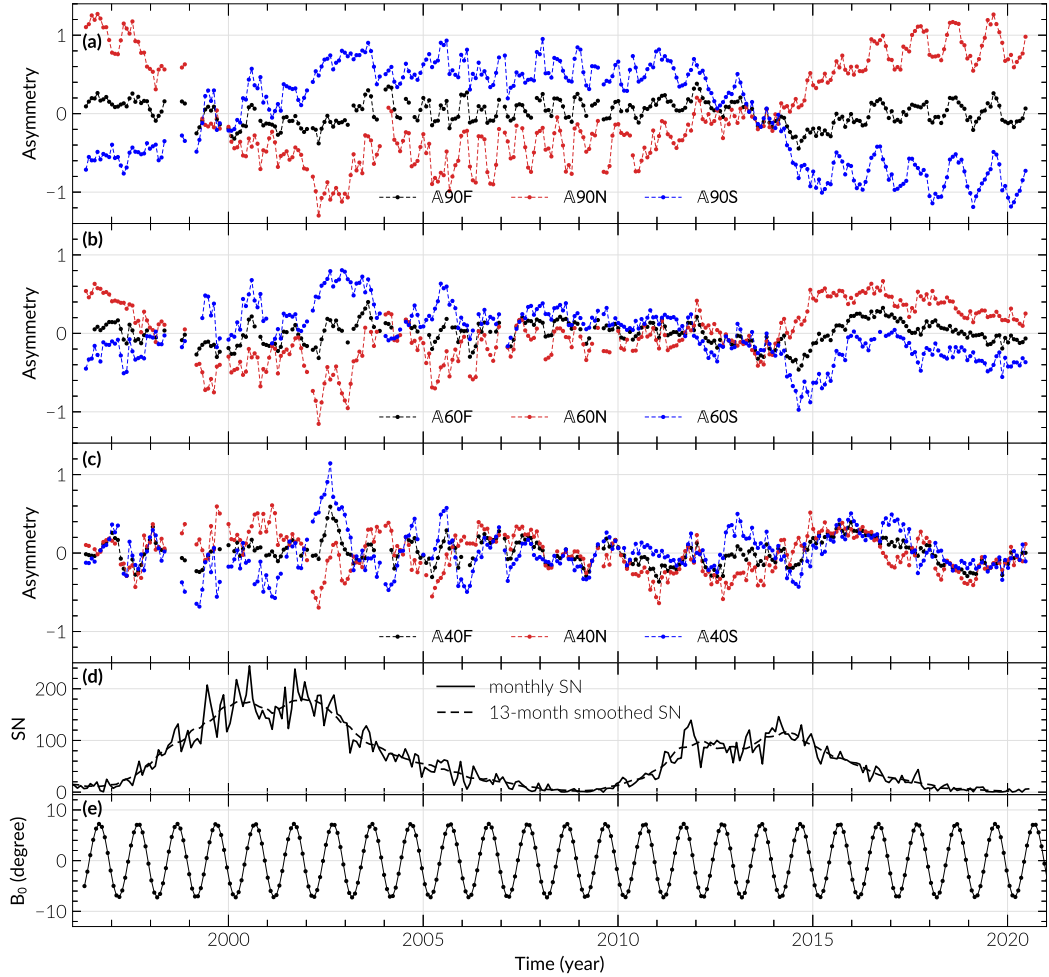}
\caption{Asymmetry values in various cases. (a) 90F (black), 90N (red), 90S (blue). (b) and (c) are similar to (a), only that the latitudes are restricted to $\pm60$\textdegree\ and $\pm40$\textdegree, respectively (see text for details). Only values above the 95\% confidence level are shown. (d) The SIDC monthly sunspot numbers (solid line) and 13-month smoothed sunspot numbers (dashed line). (e) The inclination of the solar rotational axis (B$_0$ angle) with respect to a normal to the ecliptic plane.}
\label{fig:AsymHemi}
\end{figure*}

It is of interest to exclude the high latitudes and polar regions and study the asymmetry; there are two reasons. First, the Sun's rotation axis is tilted by about 7.25 degrees with respect to the ecliptic plane, and the region near each one of the solar poles is unable to be observed during half of the year.  So either in the north or the south, some pixels in the polar regions in synoptic charts are absent of values.
Second, in synoptic charts the high latitude regions are substantially exaggerated in the toroidal direction, which is due to the method of construction of synoptic charts, i.e., the process of remapping and interpolating onto a Carrington coordinate grid~\citep{Ulrich2006}. This problem is not taken into consideration in previous studies.
Thus, we studied the \A\ in latitudes within $\pm60$\textdegree\ and within $\pm40$\textdegree\ as well, that is, excluding the magnetic field values of high latitudes in synoptic maps.
Furthermore, to investigate the difference of distribution of magnetic field values between the two hemispheres, we also studied the asymmetry of the northern hemisphere (N-hemisphere) and southern hemisphere (S-hemisphere) individually.

The results of \A\ values are displayed in Figure~\ref{fig:AsymHemi} (a), (b), and (c). Note that the asymmetry values are calculated only in case-3. In the figure, the legend of the line indicates the range of latitude that is used to calculate the asymmetry values. For example, in panel (b), `\A60F' means result of latitude from $-60$\textdegree\ to 60\textdegree (see Figure~\ref{fig:syn_chart}), `\A60N' means 0\textdegree\ to  60\textdegree\ (in the N-hemisphere), and `\A60S' means $-60$\textdegree\ to 0\textdegree\ (in the S-hemisphere).
Only \A\ values that are above the 95\% confidence level are shown with dots connected by dashed lines (some of them are not connected because there are insignificant or missing values between them).
The daily sunspot numbers and the 13-month filtered sunspot numbers are displayed in panel (d) to show the solar cycle. The variation of the B$_0$ angle in the range $\pm$7.25\textdegree\ during the contemporary time is shown in panel (e).
The main findings from the results in Figure~\ref{fig:AsymHemi} can be summed up as follows:
\begin{itemize}
\item  from panel (a) to (c), the magnitude of hemispheric \A\ decreases overall after high-latitude regions are excluded;
\item  in panel (a), the variations of hemispheric \A\ values (\A90N and \A90S) reflect the solar cycle, and show obvious flips at solar maxima;
\item  during solar minima, \A90N and \A90S differ the most and show opposite signs; on the contrary, \A40N and \A40S agrees better during solar minima;
\item  the hemispheric or full-disk \A\ in panel (a) show an obvious annual variation, while such a periodicity is not that apparent in panels (b) and (c);
\item  the magnitude of the asymmetry values calculated when both of the two hemisphere are considered  (i.e., \A90F, \A60F, and \A40F) are close.
\end{itemize}

\begin{figure*}
\centering
\includegraphics[width=1.0\textwidth]{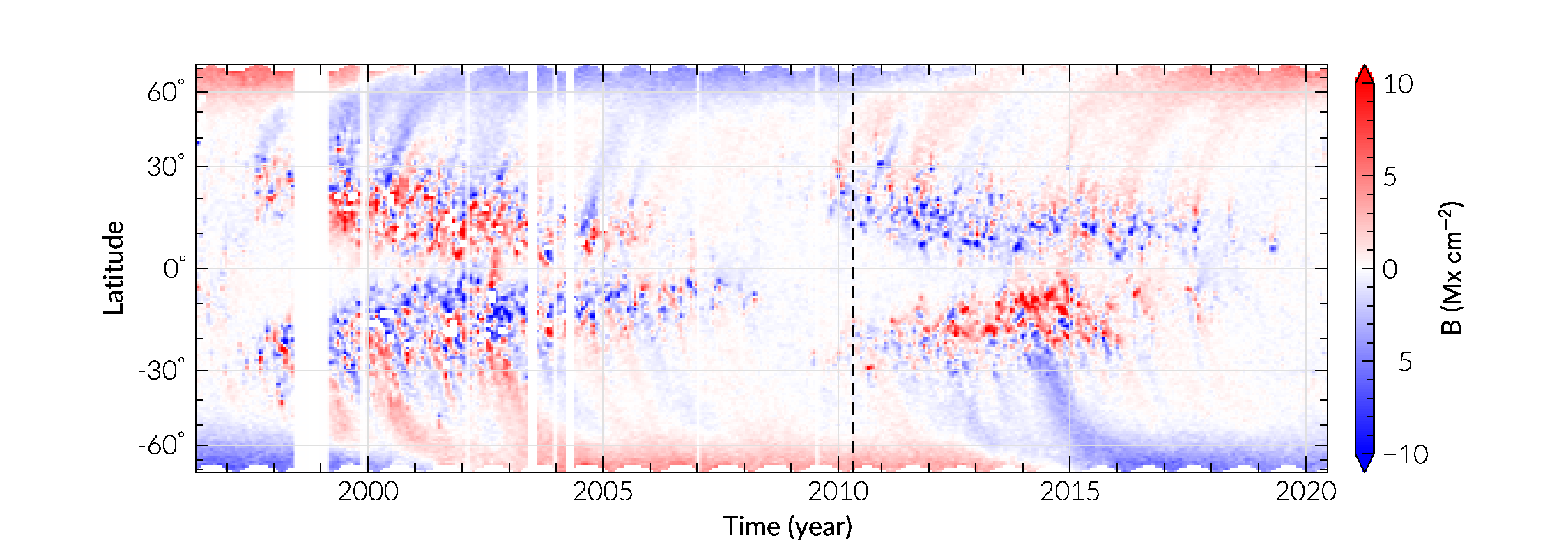}
\caption{A magnetic butterfly diagram constructed from synoptic charts of MDI and HMI (divided by the vertical dashed line). The values of MDI are scaled down by a factor of 1.4 to make the two parts more consistent. Values exceed $\pm$10 Mx cm$^{-2}$ are represented by red or blue. Blank vertical areas indicate that there are missing vales. The effect of B0 angle is clearly seen from the top and bottom wave edge.}
\label{fig:bfly}
\end{figure*}

The different results of \A\ are inferred to be caused by the emergence and evolution of magnetic flux over the entire solar disk from the equator to the poles. 
To better understand the variations of the asymmetry values, we construct a magnetic butterfly diagram during the years between 1996 and 2020 from the synoptic charts of MDI and HMI. For each chart, the magnetic field values are averaged over longitude. The diagram is shown in Figure \ref{fig:bfly}. In the figure, the values of MDI are scaled down by a factor of 1.4 to make the two parts more consistent. Values exceeding $\pm$10 Mx cm$^{-2}$ are represented by red or blue, respectively. In the left half, there are some blank vertical areas which indicate that there are missing values. At the top and bottom wave-shaped edges, the effect of B$_0$ angle can be clearly seen. The Spörer's law, Hale’s polarity law, and Joy’s law are well manifested (more discussions in the final section).  

It is clearly presented in Figure \ref{fig:bfly} that the active belt is dominated by emergent bipolar flux, while the high-latitude regions are dominated by unipolar flux surges.
When both high- and low-latitudes are considered in the calculation, the asymmetries are greatly modulated by the flux in high-latitude, as well as polar fields; with high-latitude areas excluded, the amplitude of asymmetries decreases. 
The reversal of polar fields as shown in Figure \ref{fig:bfly} is well reflected in the reversal of \A90N and \A90S during the two solar maxima.
During solar minima, the flux emergence are sparse, and \A90N and \A90S are dominated by polar region flux surges. 
The relatively small and similar amplitudes of the asymmetries \A90F, \A60F, and \A40F indicate the approximately balanced flux on the solar surface. However, even within $\pm$40\textdegree (\A40F/N/S), the positive and negative flux seems not totally balanced at all, but rather exhibits systematic deviations.

\subsection{Periodicity of the hemispheric asymmetry values}  \label{subsec:hemi_wavelet}

\begin{figure}
\centering
\includegraphics[width=0.5\textwidth]{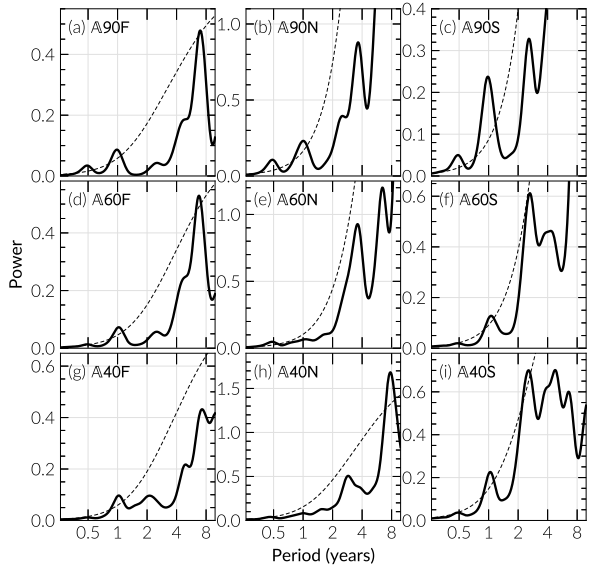}
\caption{Global wavelet power spectra of the \A\ values obtained in nine different cases as shown in Figure~\ref{fig:AsymHemi}. In each panel, the solid curve is the power spectrum and the dashed line indicates the 95\% confidence level.}
\label{fig:gwps}
\end{figure}

To study the periodicity in hemispheric asymmetry values and the \A\ that calculated after excluding high-latitude areas, we perform wavelet analyses on all of the obtained \A\ values in Figure~\ref{fig:AsymHemi} (a), (b), and (c), respectively. Their global wavelet power spectra are displayed in Figure~\ref{fig:gwps}.
Surprisingly, almost all of them show the $\sim$0.5 year and the $\sim$1 year period (above the 95\% confidence level), even when the calculation is restricted to low latitudes (i.e., \A40F, \A40N, and \A40S).

Longer periods are also shown in some of the \A\ values. Similar features are shown in each column in Figure~\ref{fig:gwps}. For instance, both \A90F and \A60F show a significant $\sim$6.7-year period (it is also shown in \A40F but below the 95\% confidence level). 
Asymmetry values of the S-hemisphere, i.e., \A90S, \A60S, and \A40S show a period of  $\sim$2.5-year; however, it is above the 95\% confidence level only in \A60S.
\A90N and \A60N show a period of $\sim$3.5 year, but all of them are not significant.
Besides, \A40N show a significant period of $\sim$7.5 years. 
The variation of this approximate solar-cycle-length period (compared to the $\sim$6.7-year period in \A90F and \A60F) is probably due to the limited length of the \A\ values.

\subsection{Polar field reversal time}

\begin{figure}[htbp]
\begin{center}
\includegraphics[width=0.48\textwidth]{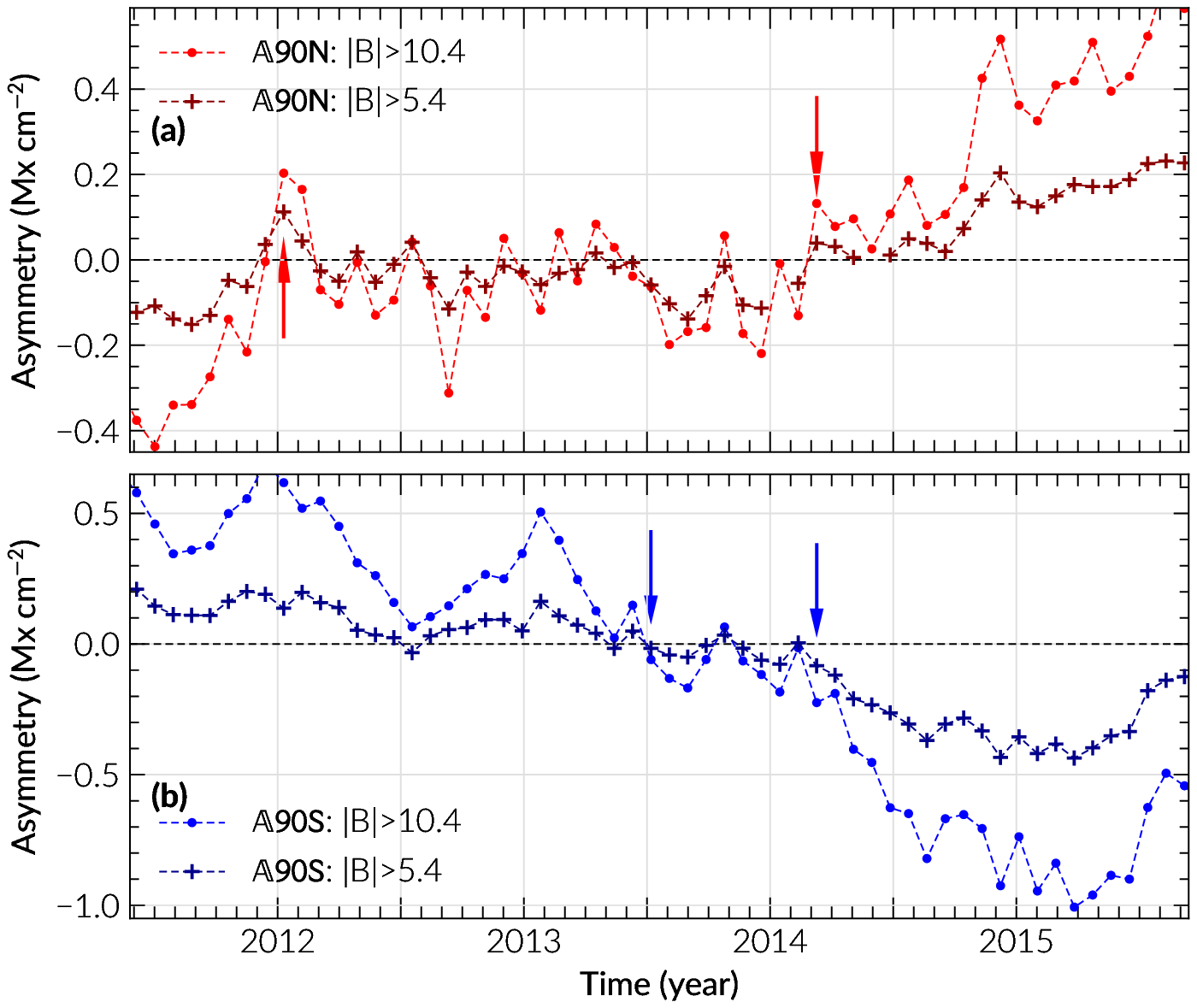}
\includegraphics[width=0.48\textwidth]{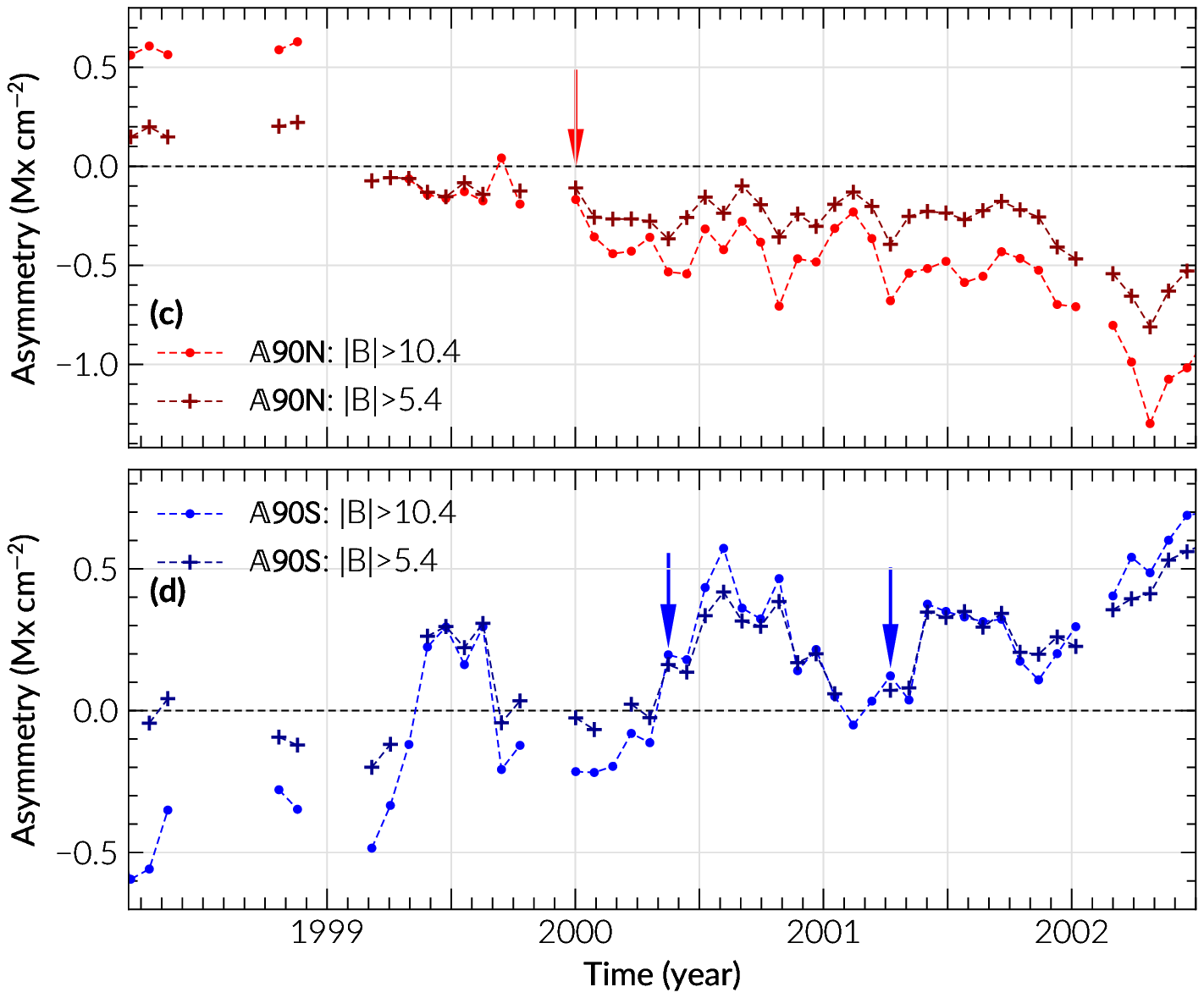}
\caption{The reversal of asymmetry values during the maximum of solar cycle 24. The upper and lower panels show the reversal of the N- and S-hemispheres, respectively. In the upper panel,  the dark-red dots display the results of case-2, and the red dots connected by dashed line display the results of case-3 (see text). The lower panel is similar to the upper panel, only that it displays the results of the S-hemisphere. The blue arrow indicates the reversal time of the southern hemispheric \A, and the right red arrow indicates that of the N-hemisphere; the left red arrow indicates a possible time that may be falsely considered a reversal.}
\label{fig:rev_time}
\end{center}
\end{figure}

In Figure~\ref{fig:AsymHemi} (a), the hemispheric asymmetry values (\A90N and \A90S) show an obvious flip at solar maxima of cycles 23 and 24, which is inferred to be caused by the reversal of the polarities of polar magnetic fields.
In order to show the reversal time more clearly, Figure~\ref{fig:rev_time} (a)--(d) display \A90N and \A90S during the maxima of these two cycles, calculated in both case-2 and case-3 (see Section~\ref{subsec:3cases} for details of the two cases). 

Since the data quality from the HMI is better (almost no insignificant or missing values), we discuss the polar reversal of cycle 24 first.
Figure~\ref{fig:rev_time} (a) and (b) show that the reversal of hemispheric \A\ of both the N- and S-hemispheres are not a simple and obvious process. 
For the S-hemisphere, according to Figure~\ref{fig:rev_time} (b), during CR-2138 (Jun 2013) and CR-2139 (July 2013) which are indicated with a blue arrow, \A90S turn from positive to mainly negative; however, taking both \A\ of case-2 and case-3 into consideration, it episodically turns back to positive in two of the Carrington rotations. Notably, the \A90S stays negative from CR-2148 (March 2014) onward. Therefore, we argue that the polar magnetic polarity reversal of the S-hemisphere happened in March 2014. It agrees well with that of \citet{Sun2015}, and it is merely 1--3 months earlier than the results of \citet{PastorYabar2015} and \citet{Gopalswamy2016}, and 4 months later than that of \citet{Janardhan2018}.

Previous results reveal that the reversal time of the N-hemisphere polar field is more diversified~\citep[e.g.,][]{Janardhan2018}, which is also reflected in the \A90N values in Figure~\ref{fig:rev_time} (a). The \A90N turns from negative to positive in as early as Jan 2012 (indicated with the left red arrow). However, from 2012 on, \A90N oscillates around 0 until March 2014 (from CR-2147 to CR-2148, indicated with the right red arrow in the figure) when \A90N turns positive and stays positive afterward. 
\citet{Karna2014}, \citet{Sun2015}, and \citet{PastorYabar2015} resulted in 2012 or early 2013 for the reversal of the N-hemisphere. \citet{Gopalswamy2016} and \citet{Janardhan2018}  gave intervals of Oct 2012--Sep 2015 and Jun 2012--Nov 2014, respectively. The intervals given by the latter two authors generally agree with the interval of complicated reversal process of \A90N shown in Figure~\ref{fig:rev_time}.
We claim that the polar magnetic polarity of the N-hemisphere reversed during the same time as that of the S-hemisphere --- in March 2014.

This result agrees well with the polar polarity changes shown in the magnetic butterfly diagram in Figure~\ref{fig:bfly}.
Specifically, the reversal time of the S-hemisphere is clearly in between 2014 and 2015, during when the polar polarity changes sign.
In the N-hemisphere, the reversal time is not that obvious, and the complication of \A\ values is explained by the magnetic field.  Although it seems like the reversal (from negative to positive) is in 2012-2013, it is affected by the obvious negative trailing polarity during 2010 and 2014. The ultimate reversal most likely happened during 2014-2015. The reversal of \A\ values confirms that it is not reversed until March 2014. 

As for solar cycle 23 shown in Figure~\ref{fig:rev_time} (d), taken both two cases of \A90S values into consideration, the polar field polarity of the S-hemisphere ultimately reversed  in March 2001 (indicated with the right blue arrow), though the first reversal happened in May 2000 as indicated with the left blue arrow. This result generally agrees with \citet{Wang2002} who stated that the S-hemisphere polarity reversal is completed in 2001.
Determining the reversal time of the N-hemisphere is impeded by a lack of data (missing or insignificant values). However, it does seems that it happened earlier than the S-hemisphere -- it might happen in Jan 2000 (or earlier) as indicated with the red arrow.

\section{Conclusion and Discussion}

In this research, we investigated the asymmetry of solar photospheric magnetic field values and its temporal variations with data of  synoptic charts spanning the recent 24 years (covering solar cycles 23 and 24). Note that the synoptic charts are constructed with magnetograms that are all zero-level offset corrected \citep{Liu2004, Liu2012}.

The distributions of magnetic field values in the synoptic charts are found to be Lorentzian rather than Gaussian. Besides an obvious better fitting at the central part of the distribution, especially, the Lorentzian function better represents the stronger magnetic fields at the tails of the distribution, which probably reflects the gradual evolution process of magnetic fields from strong fields to quiet Sun magnetic fields.

The asymmetry of distribution is uncovered by the Lorentzian function fitting, despite that it is somewhat concealed by the symmetrically distributed noise and very weak fields. Further, after excluding the noise, as well as weak fields that are indistinguishable from the noise, the asymmetry is clearly revealed. We argue that it is the stronger small-scale magnetic fields rather than supergranulations that bring about the asymmetry. 
In a resolution-reduced synoptic chart, the magnetic field value of each pixel is coming from the average of values of dozens or even thousands (depending on the resolution of the resulted chart) of pixels from the original chart; consequently, the strong magnetic fields are likely averaged out to the range within $\pm$10 G, which leads to the calculated asymmetry in previous studies \citep{Getachew2019a, Getachew2019}.
The magnetic fields of supergranulation mainly reside on the cell boundaries and thus can be resolved only with high-resolution observations \citep{Rincon2018}. Even if the asymmetry is caused by the supergranulation, it can only be detected with high-resolution observations.

We obtained asymmetry values in several circumstances (see sections \ref{subsec:3cases} and \ref{subsec:FNS}), and also studied their periodicity. Taken all of the \A\ values into consideration, almost all of them show the half-year and/or annual periods; additionally, some of them show a period approximately half of the solar cycle.
We infer that the annual (and half-year) period is caused by the variation of the B$_0$ angle, and the asymmetry is modulated by the solar cycle, i.e., the evolution of the global magnetic field. More details are discussed in the following paragraph.
These characteristics are not revealed at all in previous studies.

The various features of the asymmetry values are a result of the emergence and transportation of surface magnetic flux. 
The typical butterfly diagram (sunspot areas as a function of latitude and time, i.e., Spörer's law) manifests that active regions emerge at higher latitudes at the beginning of each cycle (within $\pm40$ degree); as the cycle progresses, the central emergent latitudes drift towards the equator. 
The emerged flux at active latitudes usually consists of bipolar active regions with a leading and a following part; the flux of these two opposite polarity parts are approximately equal.
Between the two hemispheres, the polarities of the leading (or following) parts are opposite; and also, from one solar cycle to another, of the same hemisphere, the polarities of the leading and following parts change alternatively~\citep[Hale's polarity law,][]{Hale1925}.
The leading polarity is usually more compact in morphology and intense in magnetism.
Statistically, the leading polarity spots of the emerged bipolar active regions are at a lower latitude than the following polarity; the latitudinally averaged tilt angles increase with latitude, (and the overall averaged title angle is about 5--6 degrees)~\citep[Joy's law,][]{Hale1919}.
The emerged flux is dispersed by the differential rotation, the turbulent diffusion, and the meridional flow. Both polarities of the bipolar regions are being carried poleward. 
The dispersed flux from the leading polarities in one hemisphere, since it is closer to the equator, is more likely drifting toward the equator to meet and cancel with the opposite-polarity dispersed flux of the leading polarity of the other hemisphere. 
On the contrary, the following polarities at higher latitudes have a higher probability of reaching the pole instead of being canceled. 
Also, The leading polarity, while being carried poleward, is more likely to  meet and cancel with the following polarity from the same bipole.
Some of these characteristics are well displayed in the magnetic butterfly diagram in Figure~\ref{fig:bfly} and the asymmetry values.

The calculation of the three \A40 is restricted to the activity belt ($\pm$40 degree).
Though there are lots of flux emergence during solar maxima, the emerged bipolar regions are approximately flux-balanced, and the amplitudes of \A40 are small (compared with \A90N/S). 
Surprisingly, during solar minimum, even when there are very few or no sunspots and the flux emergence is sparse, \A40N and \A40S still systematically deviate from zero, and the three generally agree well with each other. 
During times other than the solar minimum, \A40N and \A40S are anti-correlated sometimes, which could be caused by cross-equator flux cancellation events; during such processes, the total flux of leading polarities on both hemispheres decreases, and thus the asymmetries \A40 will be dominated by the following polarities \citep{Cameron2013}.
The magnitude of the \A40 variation seems independent of the solar cycle. It is not clear where does this asymmetry comes from. The magnetic elements that cause the asymmetry in \A40 seems to come from a source independent of the global solar dynamo. We speculate that it is probably related with magnetic flux coming from a local turbulent dynamo \citep[e.g.,][]{Jin2011}. 
Another probability is that the \A40 variation is caused by the cross-equatorial flux plumes \citep{Cameron2013}, as well as the variation of the B$_0$ angle. That is to say, the asymmetries brought about by cross-equatorial flux plumes are further modulated by the annual variation of the solar axis inclination, which is noteworthy even in the \A40F.

When high-latitudes and polar regions are taken into consideration, the asymmetries \A90N/S are dominated by the poleward polarity surges and polar fields. 
The polar fields are often located at $\pm$60 degrees poleward, and they are almost unipolar and relatively weak and have large spatial scales~\citep{Petrie2017}. Thus the amplitudes of \A90N/S are substantially larger.
The variation of the polar field strength lags the $\sim$22-year magnetic activity cycle by $\sim$11-year in time phase. During solar minima, the magnetic activity in the solar surface is at a low level, while the strength of polar fields reaches the peak, so the amplitudes of A90N/S also reach the peak.
Each of the poles is unable to be observed during half a year periodically, and this effect is shown in the annual variation of \A90 and in Figure~\ref{fig:bfly}.

As shown in Figure \ref{fig:AsymHemi} (a) and Figure \ref{fig:bfly}, during the late declining phase of the cycle approaching solar minima, the 0.5/1-year periodicity is most noticeable in \A90. This is because, during this time, the poleward polarity surges become less frequent and finally cease, and then the asymmetries \A90 will be dominated by high-latitude regions and polar fields. 
On the contrary, during the time when the polarity surges are strong and frequent, the 0.5/1-year modulation is partially masked by the lower-latitude streams, and thus it is not that discernible. %

We assert that the reversal of the hemispheric \A\ can be another indicator of the polar field reversal.  Especially, we show that the polar reversal of the N- and S-hemisphere of solar cycle 24 may have happened contemporarily, both in March 2014. 
It is well known that the Sun's global dipole magnetic field reverses polarity at each solar maximum during the 11-year solar cycle. Polarity reversal is essential for the Babcock-Leighton model of the solar cycle~\citep{Babcock1961, Leighton1969}. 
Since emergent active regions located in low latitudes are typically bipolar and flux-balanced while large spatial-scale high-latitude fields and polar fields are usually unipolar \citep{Petrie2017},
the hemispheric asymmetry \A90N/S mainly reflect the asymmetry of magnetic fields at high latitudes and polar regions which reverse their signs during solar maxima.
At the beginning of each cycle, the sign of the following-polarity of emergent flux is opposite to the polar fields. As the cycle progresses and reaches the maximum, the poleward surges of this flux reverse the sign of the magnetic field at high-latitude and polar regions. These surges then continue to build up new polar fields during the declining phase of the cycle\citep{Hathaway2015}.

It is a controversial issue when did the polar field of the two hemispheres reversed in solar cycles 23 and 24; different groups of researchers reported different results \citep[for a list of reversal times from previous research, see, e.g.,][]{Pishkalo2019}. 
For solar cycle 24, in previous studies, the reversal time of the N-hemisphere varies from Jun 2012 to Sep 2015, and that of the S-hemisphere varies from Nov 2013 to Jun 2014~\citep[e.g.,][]{Sun2015, Petrie2017, Janardhan2018}. 
The reversal time of the S-hemisphere determined here agrees well with that of \citet{Sun2015}.
The so-called triple or multiple reversals in the N-hemisphere in cycle 24 as indicated in, e.g., \citet{Pishkalo2019} and \citet{Janardhan2018}, is well shown in the variation of asymmetry values.
For cycle 23, the polar reversal time of the S-hemisphere determined with asymmetry values agrees with \citet{Wang2002}; furthermore, it can be assured that the reversal of the N-hemisphere was completed prior to the S-hemisphere despite that the determination of the exact date is impeded by missing values.
More research can be done to investigate the polar field reversal with the asymmetry of magnetic field values. For instance, calculating the asymmetry of distribution with daily magnetograms, restricting the calculations within high latitudes and polar regions, or calculating the asymmetry in various latitudes ranges, etc. 


\acknowledgments
Source of sunspot numbers: WDC-SILSO, Royal Observatory of Belgium, Brussels.
\href{http://soi.stanford.edu/magnetic/synoptic/carrot/M\_Corr}{SOHO/MDI} is a project of international cooperation between ESA and NASA.
\href{http://jsoc.stanford.edu/data/hmi/synoptic}{HMI} data are courtesy of the Joint Science Operations Center (JSOC) Science Data Processing team at Stanford University.
This work is supported by the National Natural Science Foundation of China (No. 11903077, 11973085, 11633008, 40636031, 11673061, 11703085),  the Specialized Research Fund for State Key Laboratories, the Basic Research Priorities Program of Yunnan (202001AU070078), and the Chinese Academy of Sciences. 

\software{Astropy (The Astropy Collaboration \citeyear{Robitaille2013, Price-Whelan2018}), Matplotlib \citep{Hunter2007}, NumPy \citep{Walt2011, Harris2020}, SciPy \citep{Virtanen2020}, IPython \citep{Perez2007}, SunPy (The SunPy Community \citeyear{Community2020}, \citealt{Mumford2020})}

\bibliographystyle{aasjournal}
\bibliography{asym_spB.bib}

\begin{thebibliography}{}
\expandafter\ifx\csname natexlab\endcsname\relax\def\natexlab#1{#1}\fi
\providecommand{\url}[1]{\href{#1}{#1}}
\providecommand{\dodoi}[1]{doi:~\href{http://doi.org/#1}{\nolinkurl{#1}}}
\providecommand{\doeprint}[1]{\href{http://ascl.net/#1}{\nolinkurl{http://ascl.net/#1}}}
\providecommand{\doarXiv}[1]{\href{https://arxiv.org/abs/#1}{\nolinkurl{https://arxiv.org/abs/#1}}}

\bibitem[{Babcock(1961)}]{Babcock1961}
Babcock, H.~W. 1961, The Astrophysical Journal, 133, 572,
  \dodoi{10.1086/147060}

\bibitem[{Bellot~Rubio \& Orozco~Su{\'a}rez(2019)}]{BellotRubio2019}
Bellot~Rubio, L., \& Orozco~Su{\'a}rez, D. 2019, Living Reviews in Solar
  Physics, 16, 1, \dodoi{10.1007/s41116-018-0017-1}

\bibitem[{Brun \& Browning(2017)}]{Brun2017}
Brun, A.~S., \& Browning, M.~K. 2017, Living Reviews in Solar Physics, 14, 4,
  \dodoi{10.1007/s41116-017-0007-8}

\bibitem[{Cameron {et~al.}(2013)Cameron, {Dasi-Espuig}, Jiang, I{\c s}{\i}k,
  Schmitt, \& Sch{\"u}ssler}]{Cameron2013}
Cameron, R.~H., {Dasi-Espuig}, M., Jiang, J., {et~al.} 2013, Astronomy \&
  Astrophysics, 557, A141, \dodoi{10.1051/0004-6361/201321981}

\bibitem[{Charbonneau(2020)}]{Charbonneau2020}
Charbonneau, P. 2020, Living Reviews in Solar Physics, 17, 4,
  \dodoi{10.1007/s41116-020-00025-6}

\bibitem[{Cheung {et~al.}(2017)Cheung, van {Driel-Gesztelyi}, Pillet, \&
  Thompson}]{Cheung2017}
Cheung, M. C.~M., van {Driel-Gesztelyi}, L., Pillet, V.~M., \& Thompson, M.~J.
  2017, Space Science Reviews, 210, 317, \dodoi{10.1007/s11214-016-0259-y}

\bibitem[{Community {et~al.}(2020)Community, Barnes, Bobra, Christe, Freij,
  Hayes, Ireland, Mumford, {Perez-Suarez}, Ryan, Shih, Chanda, Glogowski,
  Hewett, Hughitt, Hill, Hiware, Inglis, Kirk, Konge, Mason, Maloney, Murray,
  Panda, Park, Pereira, Reardon, Savage, Sip{\textbackslash}Hocz, Stansby,
  Jain, Taylor, Yadav, {Rajul}, Dang, \& Contributors}]{Community2020}
Community, T.~S., Barnes, W.~T., Bobra, M.~G., {et~al.} 2020, The Astrophysical
  Journal, 890, 68, \dodoi{10.3847/1538-4357/ab4f7a}

\bibitem[{Getachew {et~al.}(2019{\natexlab{a}})Getachew, Virtanen, \&
  Mursula}]{Getachew2019a}
Getachew, T., Virtanen, I., \& Mursula, K. 2019{\natexlab{a}}, The
  Astrophysical Journal, 874, 116, \dodoi{10.3847/1538-4357/ab0749}

\bibitem[{Getachew {et~al.}(2019{\natexlab{b}})Getachew, Virtanen, \&
  Mursula}]{Getachew2019}
---. 2019{\natexlab{b}}, Geophysical Research Letters, 0,
  \dodoi{10.1029/2019GL083339}

\bibitem[{Gopalswamy {et~al.}(2016)Gopalswamy, Yashiro, \&
  Akiyama}]{Gopalswamy2016}
Gopalswamy, N., Yashiro, S., \& Akiyama, S. 2016, The Astrophysical Journal,
  823, L15, \dodoi{10.3847/2041-8205/823/1/L15}

\bibitem[{Hale {et~al.}(1919)Hale, Ellerman, Nicholson, \& Joy}]{Hale1919}
Hale, G.~E., Ellerman, F., Nicholson, S.~B., \& Joy, A.~H. 1919, The
  Astrophysical Journal, 49, 153, \dodoi{10.1086/142452}

\bibitem[{Hale \& Nicholson(1925)}]{Hale1925}
Hale, G.~E., \& Nicholson, S.~B. 1925, The Astrophysical Journal, 62, 270,
  \dodoi{10.1086/142933}

\bibitem[{Harris {et~al.}(2020)Harris, Millman, {van der Walt}, Gommers,
  Virtanen, Cournapeau, Wieser, Taylor, Berg, Smith, Kern, Picus, Hoyer, {van
  Kerkwijk}, Brett, Haldane, {del R{\'i}o}, Wiebe, Peterson,
  {G{\'e}rard-Marchant}, Sheppard, Reddy, Weckesser, Abbasi, Gohlke, \&
  Oliphant}]{Harris2020}
Harris, C.~R., Millman, K.~J., {van der Walt}, S.~J., {et~al.} 2020, Nature,
  585, 357, \dodoi{10.1038/s41586-020-2649-2}

\bibitem[{Harvey(1971)}]{Harvey1971}
Harvey, J. 1971, Publications of the Astronomical Society of the Pacific, 83,
  539, \dodoi{10.1086/129171}

\bibitem[{Hathaway(2015)}]{Hathaway2015}
Hathaway, D.~H. 2015, Living Reviews in Solar Physics, 12,
  \dodoi{10.1007/lrsp-2015-4}

\bibitem[{Hoeksema {et~al.}(2014)Hoeksema, Liu, Hayashi, Sun, Schou, Couvidat,
  Norton, Bobra, Centeno, Leka, Barnes, \& Turmon}]{Hoeksema2014}
Hoeksema, J.~T., Liu, Y., Hayashi, K., {et~al.} 2014, Solar Physics, 289, 3483,
  \dodoi{10.1007/s11207-014-0516-8}

\bibitem[{Hunter(2007)}]{Hunter2007}
Hunter, J.~D. 2007, Computing in Science Engineering, 9, 90,
  \dodoi{10.1109/MCSE.2007.55}

\bibitem[{Janardhan {et~al.}(2018)Janardhan, Fujiki, Ingale, Bisoi, \&
  Rout}]{Janardhan2018}
Janardhan, P., Fujiki, K., Ingale, M., Bisoi, S.~K., \& Rout, D. 2018,
  Astronomy \& Astrophysics, 618, A148, \dodoi{10.1051/0004-6361/201832981}

\bibitem[{Jin {et~al.}(2011)Jin, Wang, Song, \& Zhao}]{Jin2011}
Jin, C.~L., Wang, J.~X., Song, Q., \& Zhao, H. 2011, The Astrophysical Journal,
  731, 37, \dodoi{10.1088/0004-637X/731/1/37}

\bibitem[{Karna {et~al.}(2014)Karna, Hess~Webber, \& Pesnell}]{Karna2014}
Karna, N., Hess~Webber, S.~A., \& Pesnell, W.~D. 2014, Solar Physics, 289,
  3381, \dodoi{10.1007/s11207-014-0541-7}

\bibitem[{Komm {et~al.}(2015)Komm, De~Moortel, Fan, Ilonidis, \&
  Steiner}]{Komm2015}
Komm, R., De~Moortel, I., Fan, Y., Ilonidis, S., \& Steiner, O. 2015, Space
  Science Reviews, 196, 167, \dodoi{10.1007/s11214-013-0023-5}

\bibitem[{Lagg {et~al.}(2017)Lagg, Lites, Harvey, Gosain, \&
  Centeno}]{Lagg2017}
Lagg, A., Lites, B., Harvey, J., Gosain, S., \& Centeno, R. 2017, Space Science
  Reviews, 210, 37, \dodoi{10.1007/s11214-015-0219-y}

\bibitem[{Leighton(1969)}]{Leighton1969}
Leighton, R.~B. 1969, The Astrophysical Journal, 156, 1, \dodoi{10.1086/149943}

\bibitem[{Liu {et~al.}(2004)Liu, {Xuepu Zhao}, \& Hoeksema}]{Liu2004}
Liu, Y., {Xuepu Zhao}, \& Hoeksema, J.~T. 2004, Solar Physics, 219, 39,
  \dodoi{10.1023/B:SOLA.0000021822.07430.d6}

\bibitem[{Liu {et~al.}(2012)Liu, Hoeksema, Scherrer, Schou, Couvidat, Bush,
  Duvall, Hayashi, Sun, \& Zhao}]{Liu2012}
Liu, Y., Hoeksema, J.~T., Scherrer, P.~H., {et~al.} 2012, Solar Physics, 279,
  295, \dodoi{10.1007/s11207-012-9976-x}

\bibitem[{Mumford {et~al.}(2020)Mumford, Freij, Christe, Ireland, Mayer,
  Hughitt, Shih, Ryan, Liedtke, {P{\'e}rez-Su{\'a}rez}, Chakraborty, K, Inglis,
  Pattnaik, Sip{\H o}cz, Sharma, Leonard, Stansby, Hewett, Hamilton, Hayes,
  Panda, Earnshaw, Choudhary, Kumar, Chanda, Haque, Kirk, Mueller, Konge,
  Srivastava, Jain, Bennett, Baruah, Barnes, Charlton, Maloney, Chorley,
  No~Last~Name, Modi, Mason, Rozo, Manley, Chatterjee, Evans, Malocha, Bobra,
  Ghosh, Sta{\'n}czak, De~Visscher, Verma, Agrawal, Buddhika, Sharma, Park,
  Bates, Goel, Taylor, Cetusic, No~Last~Name, Inchaurrandieta, Dacie, Dubey,
  Sharma, Bray, Rideout, Zahniy, Meszaros, Bose, Chicrala, No~Last~Name,
  Guennou, D'Avella, Williams, Ballew, Murphy, Lodha, Robitaille, Krishan,
  Hill, Eigenbrot, Mampaey, Wiedemann, Molina, Ke{\c s}kek, Habib, Letts,
  Baz{\'a}n, Arbolante, Gomillion, Kothari, Sharma, Stevens, {Price-Whelan},
  Mehrotra, Kustov, Stone, Dang, Arias, Dover, Verstringe, Kumar, Mathur,
  Babuschkin, Wimbish, {Buitrago-Casas}, Krishna, Hiware, Mangaonkar, Mendero,
  Schoentgen, Gyenge, Streicher, Mekala, Mishra, Srikanth, Jain, Yadav,
  Wilkinson, Pereira, Agrawal, No~Last~Name, No~Last~Name, \&
  Murray}]{Mumford2020}
Mumford, S., Freij, N., Christe, S., {et~al.} 2020, Journal of Open Source
  Software, 5, 1832, \dodoi{10.21105/joss.01832}

\bibitem[{Pastor~Yabar {et~al.}(2015)Pastor~Yabar, Mart{\'i}nez~Gonz{\'a}lez,
  \& Collados}]{PastorYabar2015}
Pastor~Yabar, A., Mart{\'i}nez~Gonz{\'a}lez, M.~J., \& Collados, M. 2015,
  Monthly Notices of the Royal Astronomical Society: Letters, 453, L69,
  \dodoi{10.1093/mnrasl/slv108}

\bibitem[{Perez \& Granger(2007)}]{Perez2007}
Perez, F., \& Granger, B.~E. 2007, Computing in Science \& Engineering, 9, 21,
  \dodoi{10.1109/MCSE.2007.53}

\bibitem[{Petrie \& Ettinger(2017)}]{Petrie2017}
Petrie, G., \& Ettinger, S. 2017, Space Science Reviews, 210, 77,
  \dodoi{10.1007/s11214-015-0189-0}

\bibitem[{Pishkalo(2019)}]{Pishkalo2019}
Pishkalo, M.~I. 2019, Solar Physics, 294, 137,
  \dodoi{10.1007/s11207-019-1520-9}

\bibitem[{{Price-Whelan} {et~al.}(2018){Price-Whelan}, Sip{\textbackslash}Hocz,
  G{\"u}nther, Lim, Crawford, Conseil, Shupe, Craig, Dencheva, Ginsburg,
  VanderPlas, Bradley, {P{\'e}rez-Su{\'a}rez}, de~{Val-Borro}, Aldcroft, Cruz,
  Robitaille, Tollerud, Ardelean, Babej, Bach, Bachetti, Bakanov, Bamford,
  Barentsen, Barmby, Baumbach, Berry, Biscani, Boquien, Bostroem, Bouma,
  Brammer, Bray, Breytenbach, Buddelmeijer, Burke, Calderone, Rodr{\'i}guez,
  Cara, Cardoso, Cheedella, Copin, Corrales, Crichton, D'Avella, Deil, Depagne,
  Dietrich, Donath, Droettboom, Earl, Erben, Fabbro, Ferreira, Finethy, Fox,
  Garrison, Gibbons, Goldstein, Gommers, Greco, Greenfield, Groener, Grollier,
  Hagen, Hirst, Homeier, Horton, Hosseinzadeh, Hu, Hunkeler, Ivezi{\'c}, Jain,
  Jenness, Kanarek, Kendrew, Kern, Kerzendorf, Khvalko, King, Kirkby, Kulkarni,
  Kumar, Lee, Lenz, Littlefair, Ma, Macleod, Mastropietro, McCully, Montagnac,
  Morris, Mueller, Mumford, Muna, Murphy, Nelson, Nguyen, Ninan, N{\"o}the,
  Ogaz, Oh, Parejko, Parley, Pascual, Patil, Patil, Plunkett, Prochaska,
  Rastogi, Janga, Sabater, Sakurikar, Seifert, Sherbert, {Sherwood-Taylor},
  Shih, Sick, Silbiger, Singanamalla, Singer, Sladen, Sooley, Sornarajah,
  Streicher, Teuben, Thomas, Tremblay, Turner, Terr{\'o}n, van Kerkwijk, de~la
  Vega, Watkins, Weaver, Whitmore, Woillez, Zabalza, \&
  {and}}]{Price-Whelan2018}
{Price-Whelan}, a. A.~M., Sip{\textbackslash}Hocz, B.~M., G{\"u}nther, H.~M.,
  {et~al.} 2018, The Astronomical Journal, 156, 123,
  \dodoi{10.3847/1538-3881/aabc4f}

\bibitem[{Rincon \& Rieutord(2018)}]{Rincon2018}
Rincon, F., \& Rieutord, M. 2018, Living Reviews in Solar Physics, 15, 6,
  \dodoi{10.1007/s41116-018-0013-5}

\bibitem[{Robitaille {et~al.}(2013)Robitaille, Tollerud, Greenfield,
  Droettboom, Bray, Aldcroft, Davis, Ginsburg, {Price-Whelan}, Kerzendorf,
  Conley, Crighton, Barbary, Muna, Ferguson, Grollier, Parikh, Nair,
  G{\"u}nther, Deil, Woillez, Conseil, Kramer, Turner, Singer, Fox, Weaver,
  Zabalza, Edwards, Bostroem, Burke, Casey, Crawford, Dencheva, Ely, Jenness,
  Labrie, Lim, Pierfederici, Pontzen, Ptak, Refsdal, Servillat, \&
  Streicher}]{Robitaille2013}
Robitaille, T.~P., Tollerud, E.~J., Greenfield, P., {et~al.} 2013, Astronomy \&
  Astrophysics, 558, A33, \dodoi{10.1051/0004-6361/201322068}

\bibitem[{Scherrer {et~al.}(1995)Scherrer, Bogart, Bush, Hoeksema, Kosovichev,
  Schou, Rosenberg, Springer, Tarbell, Title, Wolfson, Zayer, Akin, Carvalho,
  Chevalier, Duncan, Edwards, Katz, Levay, Lindgren, Mathur, Morrison, Pope,
  Rehse, \& Torgerson}]{Scherrer1995}
Scherrer, P., Bogart, R., Bush, R., {et~al.} 1995, SOLAR PHYSICS, 162, 129

\bibitem[{Scherrer {et~al.}(2011)Scherrer, Schou, Bush, Kosovichev, Bogart,
  Hoeksema, Liu, Jr, Zhao, Title, Schrijver, Tarbell, \&
  Tomczyk}]{Scherrer2011}
Scherrer, P.~H., Schou, J., Bush, R.~I., {et~al.} 2011, Solar Physics, 275,
  207, \dodoi{10.1007/s11207-011-9834-2}

\bibitem[{Schou {et~al.}(2011)Schou, Scherrer, Bush, Wachter, Couvidat,
  {Rabello-Soares}, Bogart, Hoeksema, Liu, Jr, Akin, Allard, Miles, Rairden,
  Shine, Tarbell, Title, Wolfson, Elmore, Norton, \& Tomczyk}]{Schou2011}
Schou, J., Scherrer, P.~H., Bush, R.~I., {et~al.} 2011, Solar Physics, 275,
  229, \dodoi{10.1007/s11207-011-9842-2}

\bibitem[{Sheeley(1966)}]{Sheeley1966}
Sheeley, Jr., N.~R. 1966, The Astrophysical Journal, 144, 723,
  \dodoi{10.1086/148651}

\bibitem[{Sun {et~al.}(2015)Sun, Hoeksema, Liu, \& Zhao}]{Sun2015}
Sun, X., Hoeksema, J.~T., Liu, Y., \& Zhao, J. 2015, The Astrophysical Journal,
  798, 114, \dodoi{10.1088/0004-637X/798/2/114}

\bibitem[{Ulrich \& Boyden(2006)}]{Ulrich2006}
Ulrich, R.~K., \& Boyden, J.~E. 2006, Solar Physics, 235, 17,
  \dodoi{10.1007/s11207-006-0041-5}

\bibitem[{van~der Walt {et~al.}(2011)van~der Walt, Colbert, \&
  Varoquaux}]{Walt2011}
van~der Walt, S., Colbert, S.~C., \& Varoquaux, G. 2011, Computing in Science
  \& Engineering, 13, 22, \dodoi{10.1109/MCSE.2011.37}

\bibitem[{Virtanen {et~al.}(2020)Virtanen, Gommers, Oliphant, Haberland, Reddy,
  Cournapeau, Burovski, Peterson, Weckesser, Bright, {van der Walt}, Brett,
  Wilson, Millman, Mayorov, Nelson, Jones, Kern, Larson, Carey, Polat, Feng,
  Moore, VanderPlas, Laxalde, Perktold, Cimrman, Henriksen, Quintero, Harris,
  Archibald, Ribeiro, Pedregosa, \& {van Mulbregt}}]{Virtanen2020}
Virtanen, P., Gommers, R., Oliphant, T.~E., {et~al.} 2020, Nature Methods, 17,
  261, \dodoi{10.1038/s41592-019-0686-2}

\bibitem[{Wang {et~al.}(2002)Wang, Sheeley, \& Andrews}]{Wang2002}
Wang, Y.-M., Sheeley, N.~R., \& Andrews, M.~D. 2002, Journal of Geophysical
  Research: Space Physics, 107, SSH 10, \dodoi{10.1029/2002JA009463}

\bibitem[{Wiegelmann {et~al.}(2014)Wiegelmann, Thalmann, \&
  Solanki}]{Wiegelmann2014}
Wiegelmann, T., Thalmann, J.~K., \& Solanki, S.~K. 2014, Astronomy and
  Astrophysics Review, 22, 1, \dodoi{10.1007/s00159-014-0078-7}

\end{thebibliography}
\end{document}